# Polariton-mediated light emission induced by electric current flow in nanostructured polyaniline

Jerzy J. Langer*, Ewelina Frąckowiak, Katarzyna Ratajczak

A. Mickiewicz University in Poznań, Faculty of Chemistry,
Laboratory for Materials Physicochemistry and Nanotechnology,
Uniwersytetu Poznańskiego 8 , PL- 61614 Poznań, Poland

Keywords: exciton-polaritons, emission, polyaniline, nanostructures

We present here a new mechanism of light emission induced by the electric current in polyaniline micro- and nanostructures. This process involves the formation of excitons, exciton-polaritons and finally an exciton-polariton condensate, leading to laser-like emission. The phenomenon can be observed in a system consisting of conducting polymer nanowires – polyaniline, which strongly absorbs light – with randomly distributed microcavities. The paper reports the results of experiments indicating possible exciton-polariton condensation (EPC) and the action of the exciton-polariton polymer laser, which is directly electrically powered.

## Introduction

Polaritons are unique quasiparticles consisting of 50% light (photon) and 50% matter (exciton), Fig. 2a, created by the coupling of photons with the excitations of the material.

Effectiveness of optical pumping for laser action has been demonstrated in many laboratories, using a number of polymers, including semi-conducting polymers [3-6]. This concerns also excitons and exciton-polaritons formation. Even the condensate of exciton-polaritons (EPC) was usually achieved in optically pumped organic materials, also at room temperature.

The electrically pumped condensation was problematic due to insufficient polariton density. This has been solved recently using single-walled carbon nanotubes (SWCNTs) as an active material in a microcavity-integrated light-emitting field-effect transistor [11]. We note this material to be similar to nanostructured polyaniline in morphology and electrical features [1, 2, 15-19].

Thus, the excitons can also be formed in polyaniline system electrically powered. Interactions of excitons and photons lead to generation of polaritons, and these, created in a large number can condensed to EPC.

## Results and discussion

### Generation the excitons

The I-V characteristics, measured for all PANI samples in the series of experiments performed currently and previously [1, 2, 17-18], show the negative differential resistance, just around the threshold voltage for the light emission (Fig. 1). This is related to the generation of excitons [1].

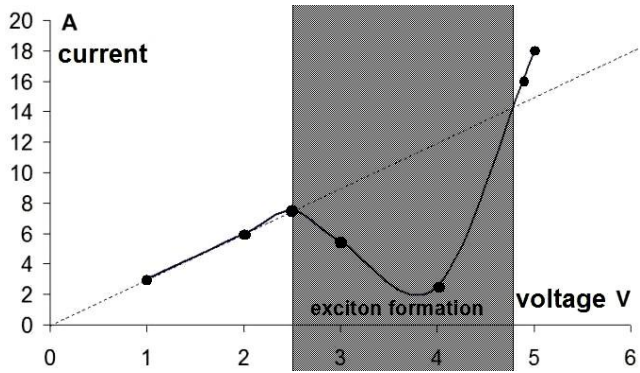

Fig. 1. I-V characteristics for PANI-$H_2SO_4$ in a whole experimental range [2]; gray area corresponds to formation the neutral excitons lowering the electrical current.

In single-wall carbon nanotubes, excitons have both Wannier-Mott and Frenkel character. This is due to the nature of the Coulomb interaction between electrons and holes in reduced-dimension. The wave function extends over a few to several nanometers along the tube axis, and the binding energy is within the range of 0.4 to 1.0 eV. Similar properties (though not identical) can be expected for excitons generated in polyaniline micro- and nanostructures.

Excitons start to be effective above a specific current, e.g. 7 A for PANI-$H_2SO_4$ (Fig. 1), which well correlates with the prediction. Assuming the life time of excitons of ~100 ps, and the Bohr radius of exciton of ~10 nm, to fill a 3 mm diameter tablet with excitons, it is necessary to generate ~$10^{20}$ photons per second, and this corresponds to the current of 5-10 A, as we found in our experiments (Fig. 1). The threshold current at 5 A is registered directly during voltage detuning experiments, when the light emission is OFF. The current when the emission is ON is much higher, reaching 22 A (Fig. 3).

**Stimulated transparency and randomly distributed optical microcavities**

Polyaniline is almost non-transparent in a normal state. It absorbs the light in a whole visual range. [17, 22] Excitons and photons generated inside a non-transparent material, are able to transform it locally into a transparent matter (a dynamic stimulated transparency owing to attenuated absorption) forming optical microcavities, in which finally polaritons are created [2, 28-30].

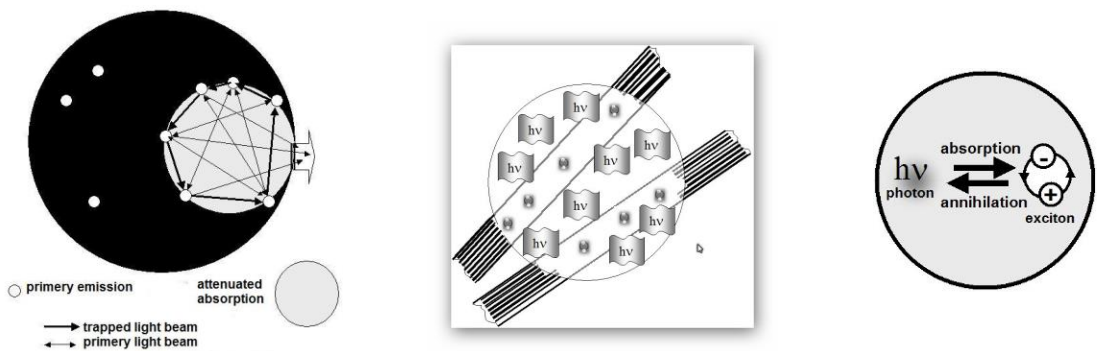

Fig. 2a. The polariton (basic concept) and a scheme of formation the polaritons in a microcavity with stimulated transparency owing to attenuated absorption in polyaniline.

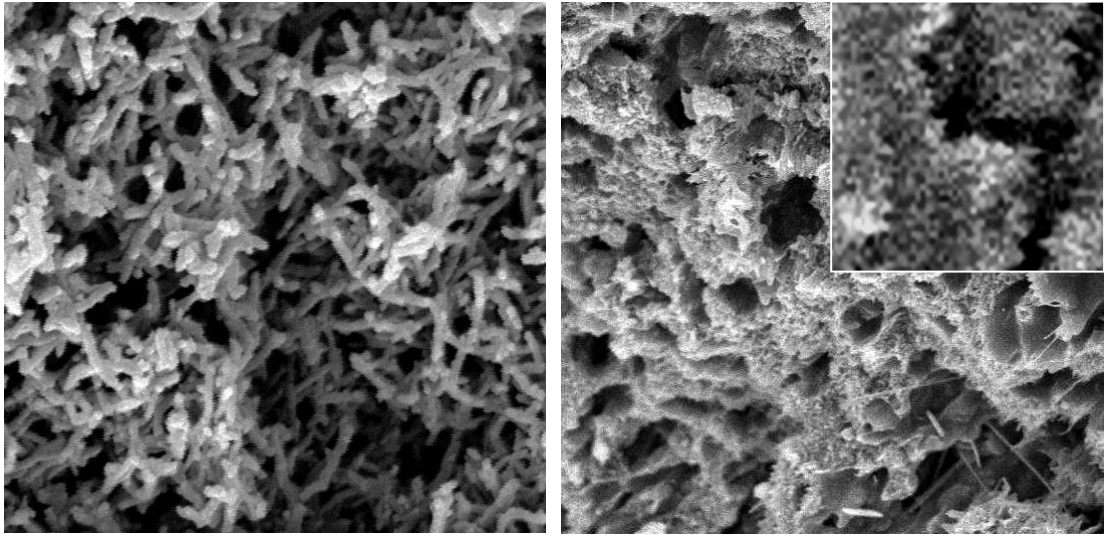

Fig. 2d. Micro- and nanostructures of polyaniline PANI-HCl (SEM, 7 μm x 7 μm; left). The cross-section of a tablet PANI-PTSA after experiment (SEM, 100 μm x 100 μm; right). The material is still micro- and nano porous (inset, SEM, 1 μm x 1 μm).

Micro- and nanostructures of the polymer (Fig. 2d), and precisely, the interactions between conducting polymer nanowires, are responsible for dynamic formation of the randomly distributed optical microcavities (Fig. 2a). This process, based on properties of polyaniline nanostructures [15,16], has been discussed in previous paper [2].

In microcavities, exciton-polaritons are created and dynamically accumulated, leading finally to a condensed system (EPC). Exciton-polaritons, as bosons of very low mass $10^{-4}$ $m_e$ ($m_e$ is electron mass), are able to do so even at room temperature [9-11, 32-33].

**Why polaritons and exciton-polariton condensate EPC are involved**

In this paper, exciton-polaritons, resulting from coupling of visible light photons with excitons, are mainly considered [8, 28-30]. Additonally, in comments to early experiments[17-18], the role of phonon-polaritons, resulting from coupling of an infrared photons with an optic phonons, is also discussed.

The active materials tested are of the same polymeric backbone which is emeraldine polyaniline. UV-VIS absorption spectrum of polyaniline film in emeraldine form shows a maximum absorption at 650 nm. This is the most probable wavelength of "polariton resonance photons" in our experiments. Thus, one can expect the polariton emission at a comparable wavelength.

In a general case, above the threshold voltage the light is polychromatic, but often includes 643 nm emission[2]. The light beam almost monochromatic red (643 nm), directional, partially collimated and coherent is observed in some samples. The emission at 643 nm is sensitive to magnetic field[8] and can be modified by a magnetic field of 200 kA/m (Fig. 2b). This is most possible the exciton-polariton condensate emission (Fig. 5).

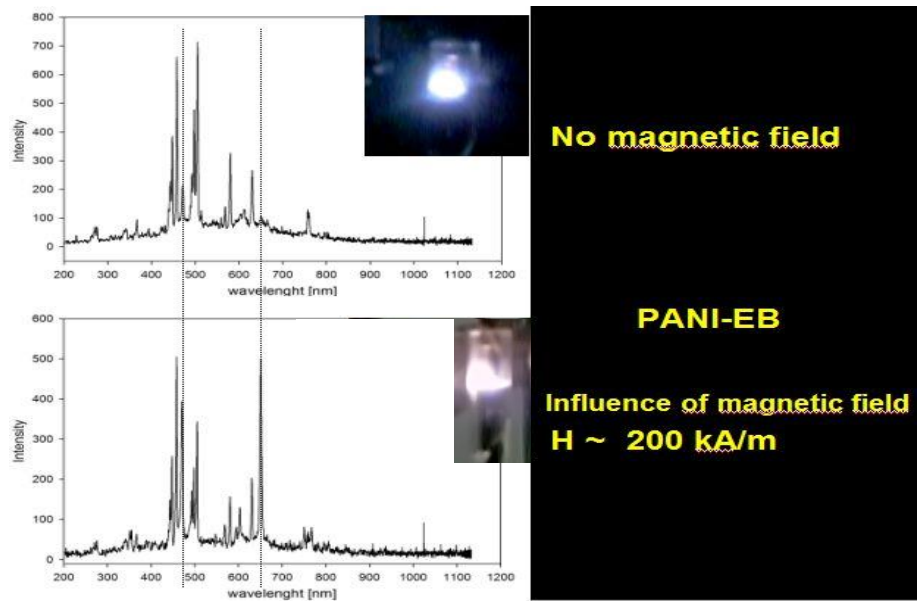

Fig. 2b. Modification of PANI-EB emission in the presence of magnetic field: a strong enhancement of the line at 643 nm is observed. The emission is clearly directional.

The emission is also sensitive to the external white light ( Fig. 2c).

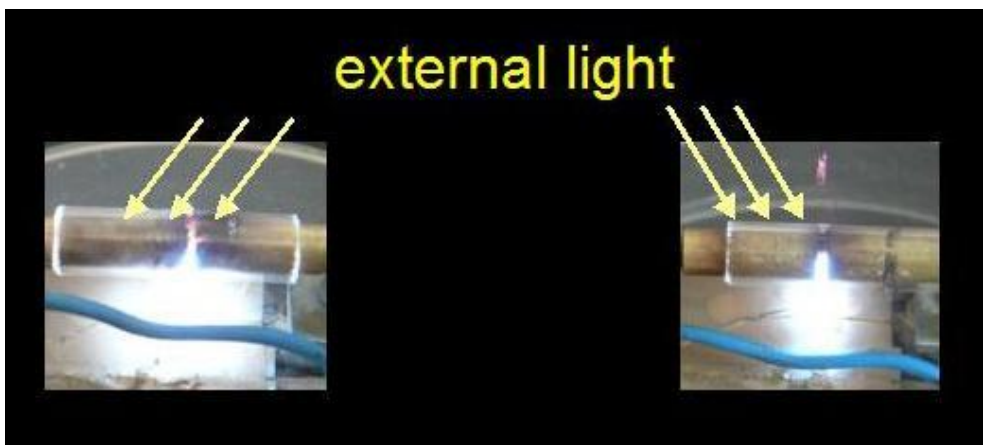

Fig. 2c. Emission is often in an opposite direction to the incident external light; this is a statistically dominant behaviour, that's why the experiments are performed in a dark room. Pictured emission from PANI-HCl and PANI- $H_2SO_4$ samples.

The results described in this paper demonstrate generation of intensive light pulses, including lasing, in nanostructured polyaniline [2, 7], doped with hydrochloric acid PANI-HCl, sulfuric acid PANI-$H_2SO_4$, p-toluenesulfonic acid PANI-PTSA, but also polyaniline emeraldine base PANI-EB, used as active materials. All the materials examined are prepared from polyaniline of the same polymeric backbone, including the emeraldine form PANI-EB.

The emission is caused by the electric current flow, above a threshold voltage, which depends on materials and experimental conditions[1,2]. The light is directional, partly collimated and can be switched reversibly ON-OFF by detuning the voltage. Switching is very sharp with the voltage changes within the range less than 1 V, while the current changes more than 15 A (5-22 A; Fig. 3).

Additionally, the wavelength and the intensity of the light are clearly dependent on the magnetic field [8] (Fig. 2b) and also on the incident external light (Fig. 2c).
Taking into account these observations, consideration should be given to polariton-mediated mechanism of the emission [8,9].

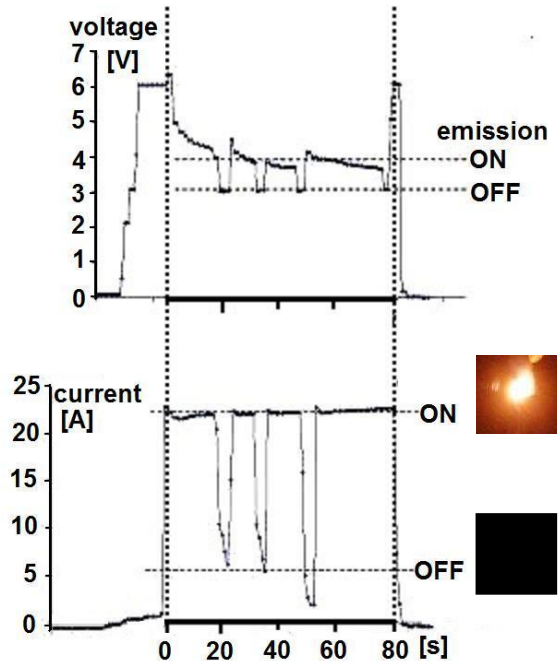

Fig. 3. Sharp switching (ON-OFF) the emission from PANI-PTSA caused by voltage changes within the range less than 1 V across the sample, while the current changes more than 15 A.

Below the threshold voltage, the emission of light is weak and proportional to the square of the current pumped, owing to the thermal generation of electrons and holes, finally - excitons. Spontaneous annihilation of excitons results in a low-intensity emission of light[2] (Fig. 4). Above the threshold voltage, the excitons are formed efficiently and the intensity of light is much higher [27]. Interactions of excitons and photons enable generation of polaritons [28-30]. When the concentration of excitons and photons is high enough, large amounts of polaritons are formed, which then condense into EPC [31]. The light emission is very intense, directional, partially collimated and linearly dependent on the pumped current. [2] (Fig. 4).

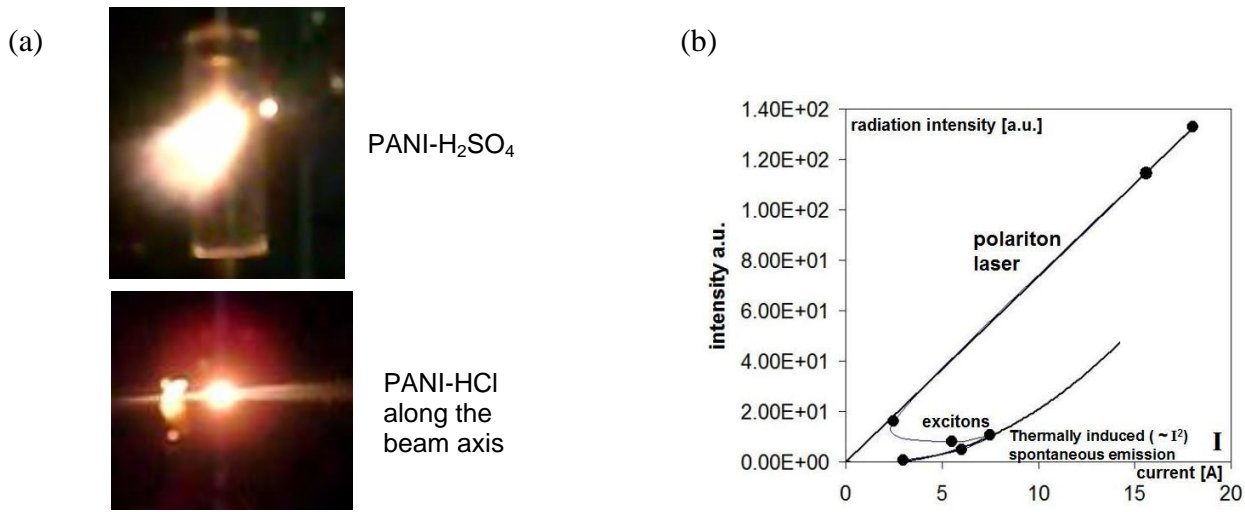

Fig. 4. Directional light emission from PANI-H$_2$SO$_4$ and PANI-HCl, registered in a dark room (a). Intensity of emission in a function of electric current pumped $^2$(b).

This is much more efficient mechanism of converting the electrical energy into the light than just a spontaneous recombination of charge carriers or annihilation of excitons. Formation of exciton-polaritons and EPC in PANI samples at ambient temperature is efficient enough at the experimental conditions applied (the high current of 20-22 A, 283-311 Acm$^{-2}$) and it don't need a high threshold energy[11, 32-33,] (here it's about 50-100 W of the electrical energy provided, maximum).
The efficiency of the electrical energy conversion into the light has been evaluated to be extremely high 80-90 %. Comparing the temperature of the sample measured bellow the threshold for emission, at the power provided of 4-8 W, and during emission at the electrical power of 75 W (Fig. 4c), one can note it is almost the same (~200 $^o$C) in both cases, and even lower during emission. Thus, most of the energy provided (80-90 % of 75 W) must be transformed into the light.

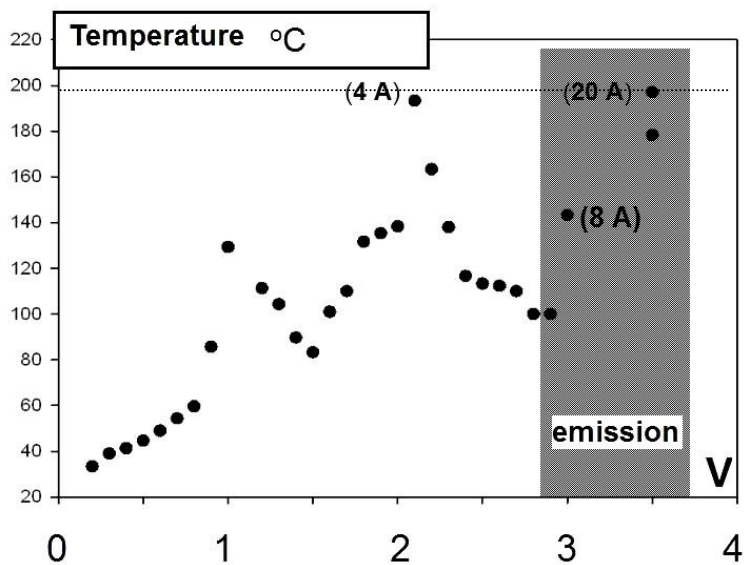

Fig. 4c. Temperature of the sample PANI-H$_2$SO$_4$ vs. the voltage applied. Please notice that, the temperature is almost the same or lower, at the current (in parentheses) of 4 A before emission and 8-20 A during emission.

The measurements are performed with the same sample at the same experimental conditions (the same heat dissipation), changing only the voltage and the current. As the temperature is almost the same (~200 °C) in both cases (even lower during emission), one can conclude, that 8 W is enough to reach the sample temperature of ~200 °C, and the excess of energy provided, i.e. 80-90 % of total electrical power, is transformed into the light. Basing on that and the emission spectrum, the extremely high efficiency of radiation about 600 lm/W is predicted. This is expected when polaritons and particularly EPC are formed, as a dynamically stable, quasi-steady state, powered by the electric current.

The EPC state combines the characteristics of lasers with those of good electrical conductors. This results in generation a light beam (which is similar to that from photon laser[33]), associated with simultaneous increase in the electrical conductivity. It is important to note, that the emission is accompanied by a series of voltage pulses at the frequency about 1 MHz, owing to periodic changes in electrical conductance of the sample, related to formation and annihilation of excitons and polaritons (Fig. 4e).

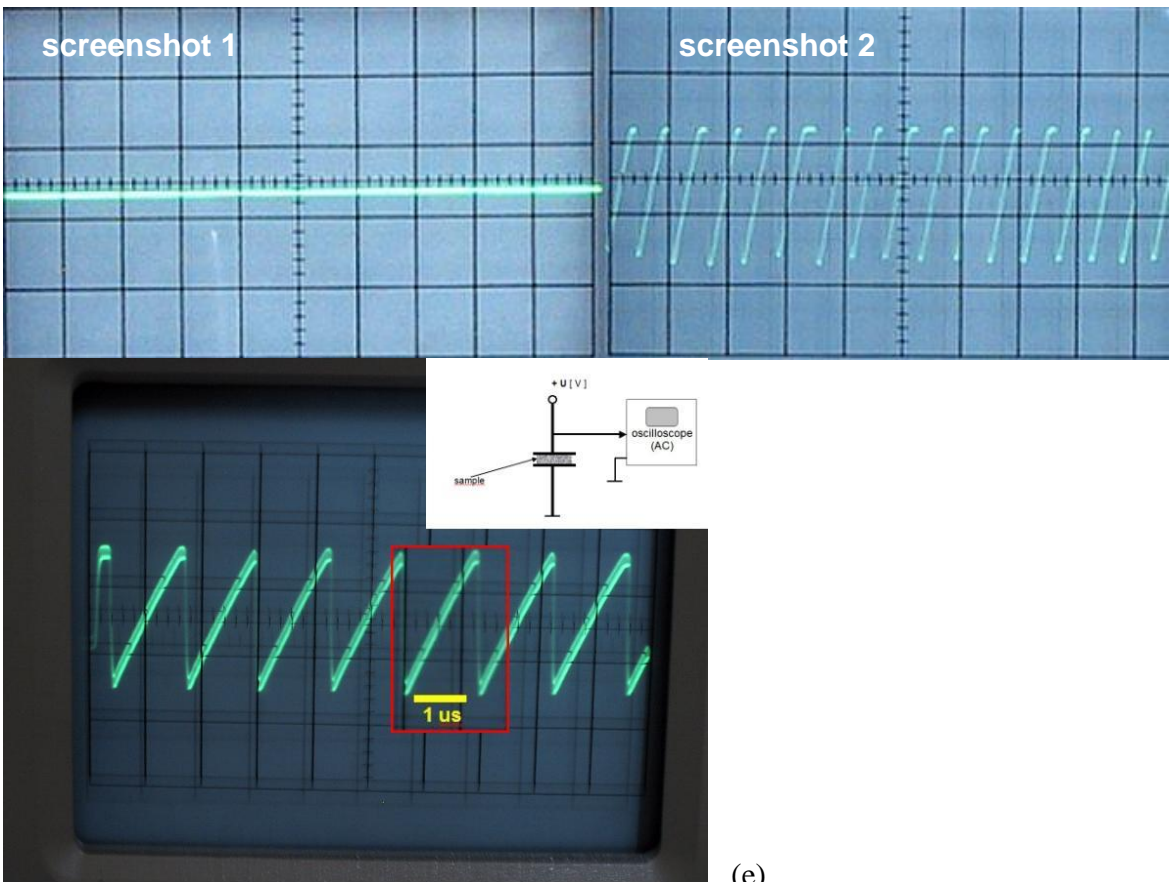

(e)

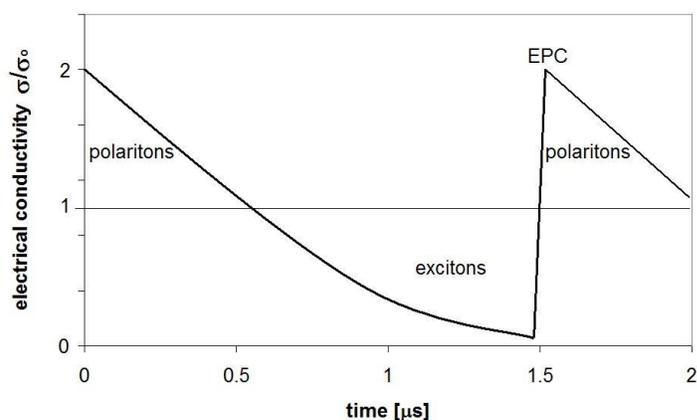

(f)

Fig. 4. Oscillations (~ 1 MHz) of the voltage at the sample during emission (e) can be assigned to formation and annihilation of excitons and exciton-polaritons - a scheme (f), where $\sigma_o$ is initial conductance of the sample (screenshot 1), and $\sigma$ is dynamic conductance during emission (screenshot 2). Condensation of polaritons (EPC) is possible at their maximum population (maximum conductance of the sample ).

Sudden changes of the resistance of polyaniline samples were also reported in previous papers, when the electrical conductivity was measured [17-21]. We found a correlation between changes in the electrical conductivity (including sharp pulses) and the IR absorption of PANI samples[17]. This demonstrates the influence of phonons onto the electrical conductivity and the results have been described as a possible superconductivity at room temperature ($T_c = 22.5\ ^oC$) [17]. This can be considered as the result of polariton condensation (EPC). The emission of light was observed during first experiments in 1975-76 [17]. Only recently has intense infrared (IR) emission been detected. On the other hand, we observe the Raman light scattering on phonon polaritons (Fig. 5a, b).

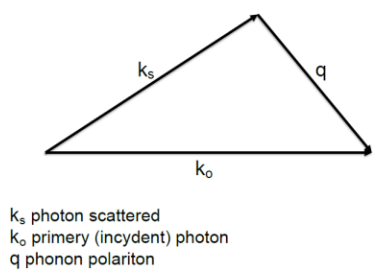

$k_s$ photon scattered
$k_o$ primery (incydent) photon
q phonon polariton

Fig. 5a. A scheme of Raman light scattering in the presence of phonon polaritons.

**Polaritons and Raman scattering**

When the emission spectrum is broad, it consists of components clearly associated with exchange of phonon energy, assigned to typical molecular vibrations in polyaniline and its salts (e.g. PANI-$H_2SO_4$ ; 1592 $cm^{-1}$, 1563 $cm^{-1}$, 1485 $cm^{-1}$, 1481 $cm^{-1}$, Fig. 5b) [2, 22, 25-26]. There is a very good correlation of energy levels of the system and the energy of phonons involved, which are related to basic molecular vibrations of polyaniline (Fig. 5b). The basic electronic structure is close to that of a polaron-reach polyemeraldine salt with the energy gape of 1.4-1.5 eV [23-24], which is 1.45 eV in our case (Fig. 5b). All energy levels are located in the range of 0-2.65 eV, i.e. within the energy gap for PANI-$H_2SO_4$ and close to the value of 2.7 eV for polyemeraldine salt [2, 23].

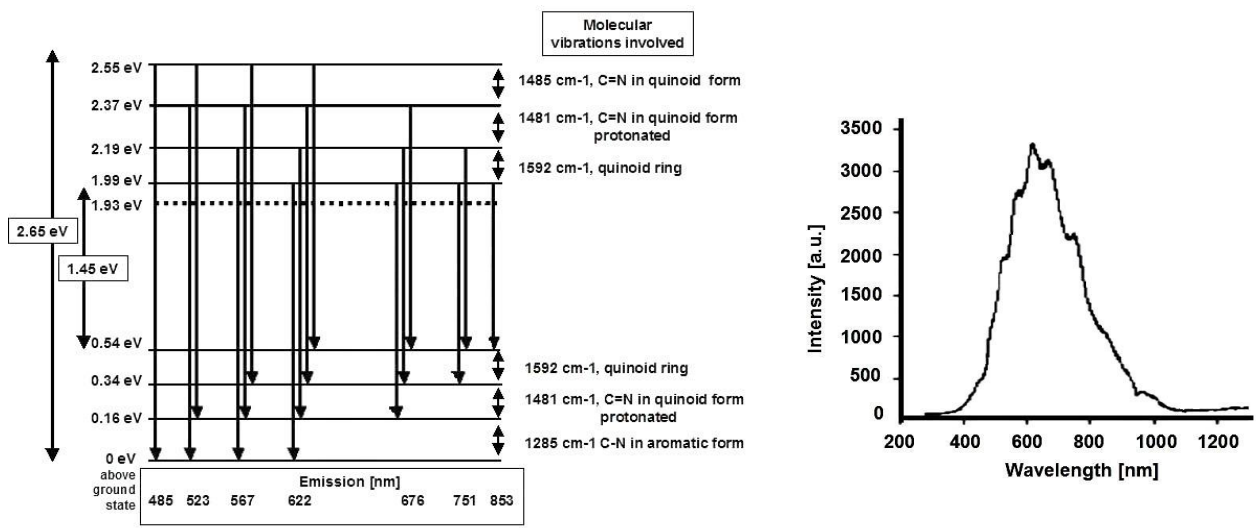

Fig. 5b. A simplified scheme of energy levels for the broad emission of PANI-$H_2SO_4$, including corresponding spectrum. Correlation of photon energy and the energy of phonons involved, which are related to basic molecular vibrations in polyaniline molecules. Photon energy level 1.93 eV corresponding to EPC emission (Fig. 5e), is marked.

**Polariton lasing**

In a part of experiments, nearly monochromatic light at 643 nm is generated (Fig. 5c), which can be assigned to EPC emission. The intensity of the line 643 nm increases while other lines are quenched at the same time (Fig. 5d). The system transforms into EPC, and in a final stage it can be described by a reduced number of quantum parameters, leading to single emission line at 643 nm. Finally, EPC is destroyed because of redundant electrical energy provided (the current cannot be too high, and this results in limited stability of emission and the current pulses observed, Fig. 4e).

The most intensive emission is observed for a single line at 643 nm. According to the formula $W_g = hc/\lambda$ (where h is Planck constant, c speed of light, and $\lambda$ wavelength of emitted light), this corresponds to the energy gap $W_g$ of 1.93 eV. In this case, there is a limited energy dissipation other than radiation, the current flows with a minimum resistance and the intensity of emission is proportional to the current (Fig. 4b). The emission of the intensive light is always associated with sudden increase in electrical conductivity.

The light intensity is proportional to the electrical energy pumped (Fig. 4b), however the energy supplied is not equally distributed among possible competitive processes other than emission at 643 nm (1.93 eV), which dominates expense others (Fig. 5d). The intensity of other emission lines is lower, and their wavelength well correlates with the energy level scheme (Fig. 5b), owing to interactions with polyaniline electronic system and molecular vibrations: 2903 $cm^{-1}$, 2626 $cm^{-1}$, 2337 $cm^{-1}$ and 2169 $cm^{-1}$ at the ground state, and 2532 $cm^{-1}$, 1073 $cm^{-1}$, 632 $cm^{-1}$, 432 $cm^{-1}$, and 415 $cm^{-1}$ when excited, where the range of 2169-2903 $cm^{-1}$ corresponds to the energy of PANI electron system, and 415-1073 $cm^{-1}$ are specific molecular vibrations.

When the emission at 643 nm is generated, a narrowing of the spectral line is observed, together with the reduction in the intensity of other lines[2], leading to limiting the number of lines, even strictly to one single narrow line of a very high intensity (Fig.5c,f).Thus, emission at 643 nm (a single narrow line), can be attributed to laser-like radiation from the system composed of strongly coupled polyaniline nanowires, in which polaritons are created and then condensed (EPC).

Recently, the electrically induced formation of polaritons in carbon nanostructured material composed of carbon nanotubes (SWCNT), which is similar to nanostructured PANI [1,2], was evidenced.[11] Previously, the electrically pumped polariton laser, based on strongly coupled quantum well microcavities, was described in 2013 by Schneider et al.[10]. Strong coupling in nanosystems was also reported by Agarwal group[27], and others, particularly in organic materials[28].

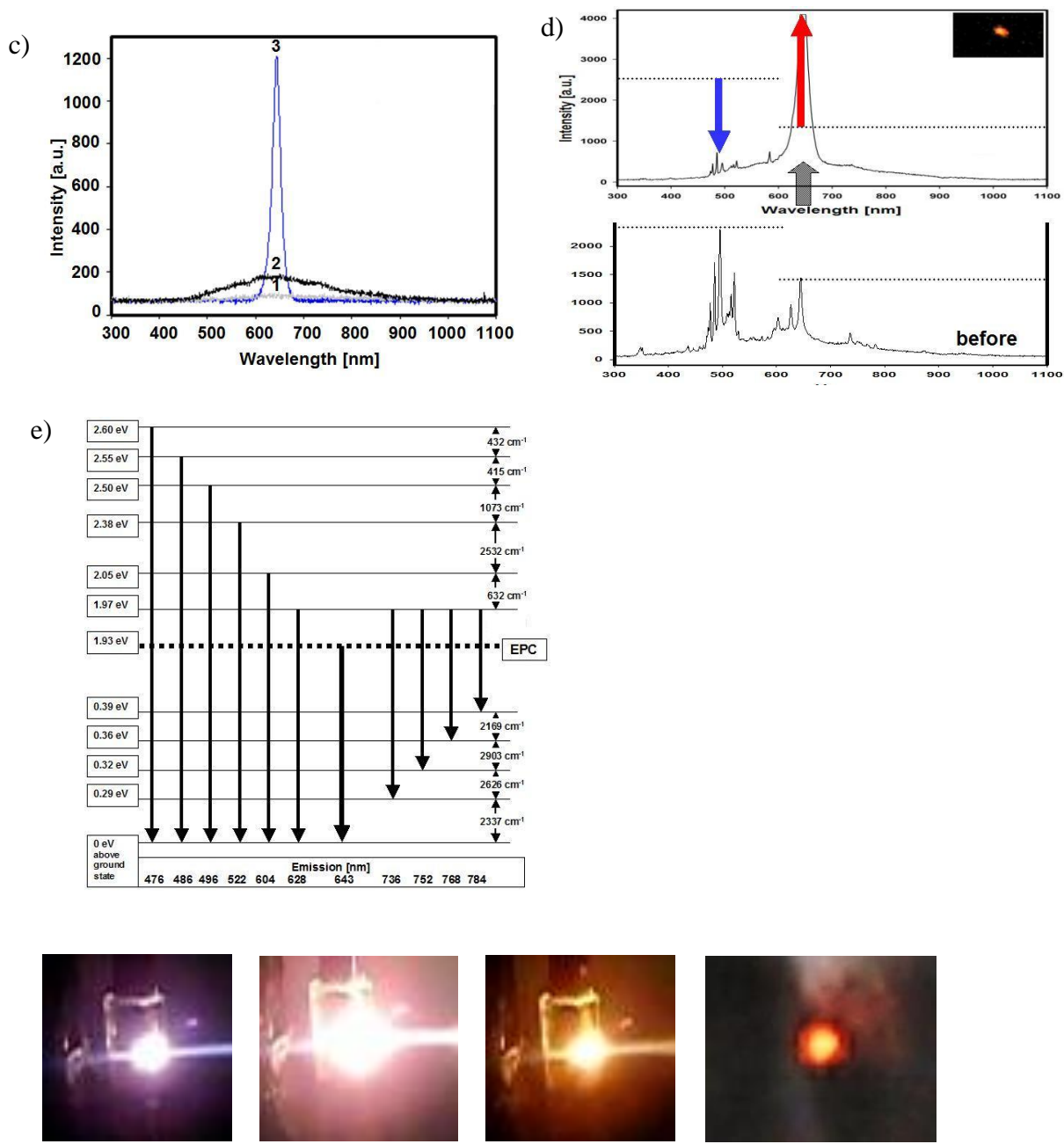

Fig. 5. Generating a single nearly monochromatic line at 634 nm in two different PANI samples: PANI-$H_2SO_4$ at 5 V, 20 A (c) and PANI-NC [2] at 14.5 V, ~5 A (d); a simplified scheme of adequate energy levels; EPC energy level corresponding to directional emission at 643 nm is shown. (e). In both cases, the intensity of the line 643 nm dramatically increases while other lines are quenched (1, 2, 3 are subsequent spectra). Light emission transformation with increased content of red line 643 nm (f).

The observed effects are summarized in Fig. 6, showing their dynamic nature due to the simultaneous multiple condensation, mobility and limited lifetime of the generated polaritons.

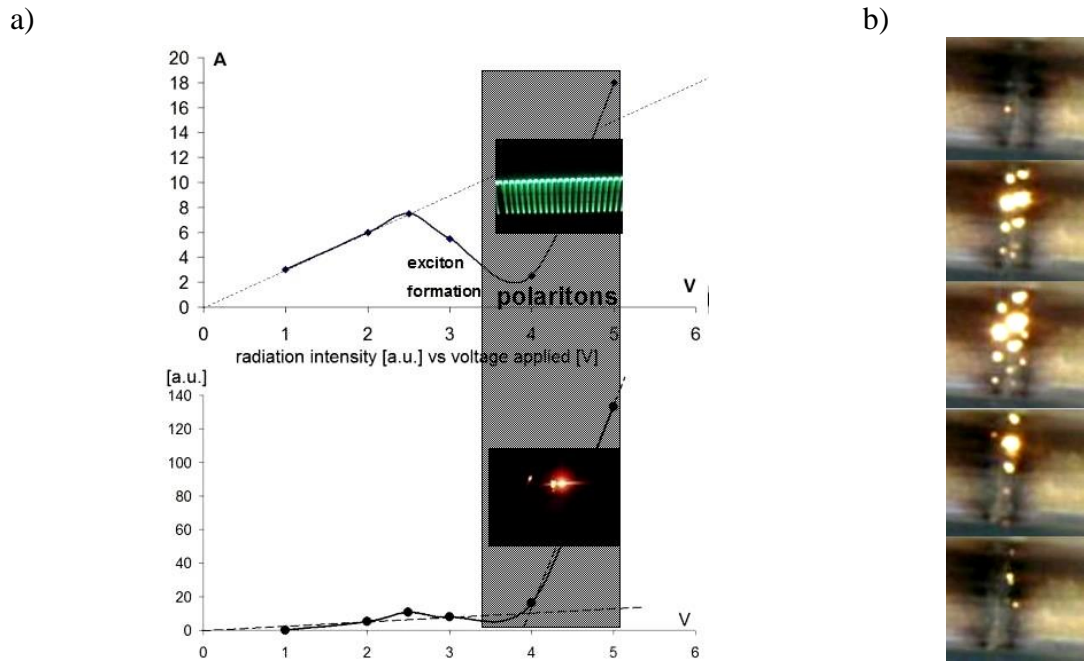

Fig. 6. Correlation of the I-V characteristics, the related emission intensity[2] and 1 MHz oscillations, induced in PANI-$H_2SO_4$ (a). Evolution of emission of PANI-HCl in time (~5 s, up-down, thickness about 1 mm) demonstrates dynamic multiple condensation owing to motility of polaritons generated (b).

**Experimental**

**Approach**

Nanostructured polyanilines PANI-EB, PANI-HCl, PANI-$H_2SO_4$ and PANI-PTSA have been prepared by oxidation of aniline hydrochloride with chemical methods, the same as described elsewhere [1, 2, 7, 12-17,22]. Micro- and nanofibrils, made of polyaniline-PS-PEO composite, were fabricated with the use of electro spinning technique, operating at the voltage of 4-6 kV at ambient conditions with controlled temperature (20-25 $^oC$) and humidity about 75 %. The resulting material PANI-NC composed of micro- and nanofibrils was treated and used the same way as other samples. All materials prepared (polyaniline and polyaniline composite micro- and nanofibrils) were characterised with physical and chemical methods (FTIR, EPR, elemental analysis) giving results identical or very similar to the values measured for materials previously prepared in our laboratory [1, 2, 15-17]. Morphology of materials examined was monitored using a scanning electron microscope (ZEISS EVO 40 or The Phenom proX).

The samples for electro luminescence studies were fabricated with micro- and nanostructured conducting polymers: PANI-EB, PANI-HCl, PANI-$H_2SO_4$ and PANI-PTSA or micro- and nanofibrils made of polyaniline composite PANI-NC, as pellets of the thickness of 0.1-0.5 mm and the diameter of 3 mm, formed at the pressure of 6000 kG/$cm^2$. These were placed in a glass tube of the wall thickness of 1-2 mm, between two solid metal electrodes (copper) of the diameter 4.5 mm and the length of 25 mm. The emission spectra were registered with the UV-VIS spectrometer Ocean Optics PC2000 at a resolution of 0.5 nm, at the same time, when the current and the voltage were measured. The beam profile was pictured directly with Olympus C-765 UZ digital camera [2] or Pentax Kr 12.4 MP +.

**Materials and methods**

Polyaniline. 5 mL of aniline was dissolved in 300 mL of 1M HCl and kept at 0$^oC$. 11,4 g of $(NH_4)_2S_2O_8$ was dissolved in 200 mL of 1M HCl and also kept at 0$^oC$. The $(NH_4)_2S_2O_8$/HCl solution was added to the aniline/HCl solution drop by drop under constant stirring over about 20 minutes. The resulting dark green solution was maintained under constant stirring for 24 hours, filtered and washed with water. The product when dried is **PANI-HCl** [1].

Wet PANI-HCl was added to 500 mL of 1M $(NH_4)OH$ and maintained under constant stirring for 24 hours. The resulting emeraldine base (**PANI-EB**) was filtered and dried.

The emeraldine base (PANI-EB) was "doped" by sulphuric acid: 500 mg of PANI-EB was added gradually to 100 mL of concentrated sulphuric acid and maintained under constant stirring for 24 hours. Then, the solution was filtered and slowly dropped to 400 mL of distilled water. After cooling, the suspension was filtered and washed with water, to get pH 4,5. Then, the product was washed with methanol to remove water. The final product polyaniline doped with $H_2SO_4$ (**PANI-$H_2SO_4$**) was separated (by filtering or centrifuging) and dried.

In the beaker, 1.4 g of p-toluenesulfonic acid was dissolved in 75 ml of chloroform. Next, 0.9 g of PANI-EB was added to this solution in small portions. The mixture was stirred with a magnetic stirrer for two hours. After two hours, the dispersion was dripped into a beaker containing 200 ml of distilled water while stirring all the time. Next, the formed precipitate was isolated by filtration under reduced pressure and washed several times with distilled water until a pH about 4,5 (checked with indicator paper) has been obtained. In order to remove most water from the solid, it was washed it several times with methanol. Finally, the obtained polyaniline protonated with p-toluenesulfonic acid (**PANI-PTSA**) was dried for 24 hours at room temperature.

Micro- and nanofibrils, made of polyaniline composite (**PANI-NC**), were fabricated with the use of electro spinning technique, operating at the voltage of 4-6 kV at ambient conditions with controlled temperature (20-25 $^oC$) and humidity about 75 %. The flow rate of basic solution was 2 mL/min. The basic solution was composed of polystyrene (PS), polyethylene oxide (PEO) and polyaniline, dissolved or suspended in chloroform at the proportion 3:1:4 (e.g. 30 mg PS, 10 mg PEO, 40 mg PANI-$H_2SO_4$ and 10 mL $CHCl_3$). Dispersion of polyaniline in chloroform (e.g. 40 mg PANI-$H_2SO_4$, 5 mL $CHCl_3$) has been prepared before with the aid of ultrasound using Ultrasonic Disintegrator UD-20 automatic (TECHPAN, Poland).

The resulting material composed of micro- and nanofibrils was treated and used the same way as other nanostructured polyaniline samples.

All materials prepared (polyaniline and polyaniline composite fibrils) were characterized with physical and chemical methods (FTIR, EPR, elemental analysis) giving results identical or very similar to the values measured for materials previously prepared in our laboratory and already published [1, 2, 15-17].

Morphology of all final materials (as described above) was monitored using a scanning electron microscope (ZEISS EVO 40 or The Phenom proX).

The samples for electro luminescence studies were fabricated with micro- and nanostructured conducting polymers: PANI-EB, PANI-HCl, PANI-$H_2SO_4$ and PANI-PTSA or micro- and nanofibrils made of polyaniline composite PANI-NC, as pellets of the thickness of 0.1-0.5 mm and the diameter of 3 mm, formed at the pressure of 6000 kG/$cm^2$, which were placed in a glass tube of the wall thickness of 1-2 mm, between two solid metal electrodes (copper) of the diameter 4.5 mm and the length of 25 mm. [2]

To register simultaneously the light beam and the emission spectrum, the optical fibre of spectrometer was mounted at the same side as photo camera, in most cases parallel to the camera optical axis, but other configuration was also used [2]. The distance between the sample and the aperture of optical fibre was of 1-3 cm, and the camera was located at a distance of 15-20 cm. [2]

The emission spectra were registered with the UV-VIS spectrometer Ocean Optics PC2000 at the resolution of 0.5 nm, at the same time, when the current and the voltage were measured. Changes in the intensity of radiation were estimated from the amplitude or the integral intensity of the most intensive spectral lines. The light beam profile was registered as a picture or a film at the framerate of 15/s, with the aid of Olympus C-765 UZ digital camera (4 Mpx) and Pentax Kr 12.4 MP + 18-55 mm lens. All the measurements were performed in a dark room at ambient conditions.

Stabilised power supply units Z-3020 (0-30 V, 20 A) INCO Poland, was used. The voltage was changed slowly (0.1-1 V/step) up to a sudden increase of current, when the light emission was observed – usually within the range 2-20 V. The voltage and the current were measured with an accuracy of 0.1 V and 0.1 A, using Brymen digital multimeter, model BM859s and Metrahit Energy Multimeter equipped with precision resistor 0,001Ω, respectively. The dynamics of current flow (pulses) was monitored with the use of Digital Waveform Recorder type 261 (UNIPAN, Poland).

Physicochemical characteristics of materials used:

The electrical conductivity of bulk samples of the nanostructured polymer (e.g. PANI-$H_2SO_4$) changes from $10^{-1}$ S/cm for a starting material to $10^0$ S/cm when activated (emission), with a negative differential conductance (typical for highly-doped semiconductors) at 3-4 V (about 300 V/cm) and a maximum differential conductance of $10^1$ S/cm at 5 V (about 500 V/cm).

The electrical conductivity of PANI-NC is lower and changes from $3.3 \times 10^{-3}$ S/cm when activated, with a negative differential conductance within the range of 8-14 V (800-1400 V/cm) and a maximum conductance of $1.3 \times 10^{-2}$ S/cm above 14 V. [2]

Chemical analyses were made using a model Vario EL III elemental analyzer (Elementar Analysensysteme GmbH, Germany).

Nanostructured PANI-$H_2SO_4$ [2]
Chemical analysis (EA): %C=51,42, %H=4,13, %N=10,31, %S=8,28, %O=25,86,
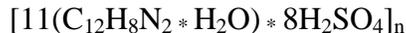
$[11(C_{12}H_8N_2 * H_2O) * 8H_2SO_4]_n$
FT-IR [cm$^{-1}$]:	2563, 1571, 1481, 1306, 1239, 1106, 982, 884, 804, 613, 508
EPR:	$\Delta H = 0.395$ mT, g = 2.0027 symmetric signal

The data correspond to polyaniline in emeraldine form partially protonated (approximately 4 $H_2SO_4$ molecules per 11 monomeric units $C_6H_4N$).

Nanocomposite PANI-NC of PANI-$H_2SO_4$, PEO and PS [2]
FT-IR [cm$^{-1}$]:	2563, 1563, 1481, 1299, 1245, 1103, 982, 879, 816, 799, 614, 504, 413
EPR:	$\Delta H = 0.422$ mT, g = 2.0032, symmetric signal

PANI-EB
Chemical analysis (EA): %C=74,05; %H=5,01; %N=13,61; %O=7,33, $[C_{12}H_8N_2 * H_2O]_n$
FT-IR [cm$^{-1}$]:	3022, 1583, 1495, 1375, 1302, 1242, 1141, 1009, 956, 824, 801, 731, 503, 411
EPR:	$\Delta H = 0,424$ mT, g = 2,0027, symmetric signal

PANI-PTSA
Chemical analysis (EA): %C=61.61, %H=5.15, %N=8.47, %S=6.25, %O=18.53,
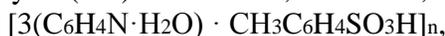
$[3(C_6H_4N \cdot H_2O) \cdot CH_3C_6H_4SO_3H]_n$,
FT-IR [cm$^{-1}$]:	3450, 1562, 1482, 1301, 1116, 1050, 1000, 802, 674, 505
EPR:	$\Delta H$ [mT] = 0.285,  g = 2.00292, asymmetry: 1.13
	$\Delta H$ [mT] = 0.236,  g = 2.00299,  asymmetry: 1.01 (after emission)

**Conclusions**

Similarity in I-V characteristics of the samples examined related to formation of excitons (despite a difference in the electrical conductivity) [2] and the identical wavelength of the radiation (643 nm) in all cases, e.g.: nanostructured PANI-EB (Fig. 2b), PANI-$H_2SO_4$ and PANI-NC [2] (Fig. 5c, d) prove, that the emission is of the same nature, and exciton-polaritons play crucial role in this process, including EPC. The conclusion is additionally supported by experiments with PANI-EB at magnetic field (Fig. 2b), high sensitivity to the voltage applied – detuning (sharp switching in PANI-PTSA, Fig. 3; such electro-optical bistability is of potential importance for future application in opto-electronics and computing [34]), sensitivity to the external light (Fig. 2c), 1 MHz oscillations (Fig. 4e, 6a) and spatial evolution of emission in time owing to motility of polaritons (Fig. 6b).

**Conflicts of interest**
There are no conflicts to declare.

**Acknowledgements**

I would like to express my heartfelt gratitude and heartfelt thanks to my beloved wife Anna for her unlimited support;
Ewelina Frąckowiak (now Ludera) and Katarzyna Ratajczak for their excellent work in our laboratory. /JJL/